\documentclass[a4paper,manyauthors,nocleardouble,COMPASS]{cernphprep}
\usepackage[utf8]{inputenc}
\usepackage{amssymb}
\usepackage{amsmath}
\usepackage{lineno}
\usepackage{xspace}
\usepackage{graphicx}
\usepackage{caption}
\usepackage{subcaption}
\usepackage{color}
\usepackage{float}
\usepackage{datetime}
\usepackage{cite}
\usepackage[numbers]{natbib}
\usepackage{hyperref}
\usepackage{amsmath,amssymb,epsfig,fleqn,url,psboxit,color,here,graphics,graphicx,rotating,multirow,wrapfig}
\usepackage{pbox}
\usepackage{dcolumn}
\usepackage{bigstrut}
\usepackage{soul,xcolor}
\usepackage{lineno, xcolor}

\usepackage[section]{placeins}

\hypersetup{
    colorlinks,
    citecolor=blue,
    filecolor=blue,
    linkcolor=blue,
    urlcolor=blue
}

\newcommand{\lf}{\left}
\newcommand{\rg}{\right}

\newcommand{\phiCS}{\varphi _{CS}}
\newcommand{\phiS}{\varphi _S}
\newcommand{\thCS}{\theta _{CS}}
\newcommand{\gvc}{GeV/$c$}

\newcommand{\gvcw}{GeV/$c^2$}

\setstcolor{red}
\begin{document}
\begin{titlepage}
\PHnumber{2017--xxx}
\PHdate{31 March January 2017}

\title{First measurement of transverse-spin-dependent azimuthal asymmetries\\ in the Drell-Yan process}

\begin{abstract}
The first measurement of transverse-spin-dependent azimuthal asymmetries in the pion-induced Drell-Yan (DY) process is reported. We use the CERN SPS 190 \gvc\, $\pi^{-}$ beam and a transversely polarized ammonia target. Three azimuthal asymmetries giving access to different transverse-momentum-dependent (TMD) parton distribution functions (PDFs) are extracted using dimuon events with invariant mass between 4.3~\gvcw\, and 8.5~\gvcw.
The observed sign of the Sivers asymmetry is found to be consistent with the fundamental prediction of Quantum Chromodynamics (QCD) that the Sivers TMD PDFs extracted from DY have a sign opposite to the one extracted from semi-inclusive deep-inelastic scattering (SIDIS) data.
We present two other asymmetries originating from the pion Boer-Mulders TMD PDFs convoluted with either the nucleon transversity or pretzelosity TMD PDFs.
These DY results are obtained at a hard scale comparable to that of a recent COMPASS SIDIS measurement and hence allow unique tests of fundamental QCD universality predictions.


\end{abstract}

\vspace*{60pt}

\vfill
\Submitted{(to be submitted to Phys.\ Rev. Lett.)}

%
%
%
\section*{The COMPASS Collaboration}
\label{app:collab}
\renewcommand\labelenumi{\textsuperscript{\theenumi}~}
\renewcommand\theenumi{\arabic{enumi}}
\begin{flushleft}
M.~Aghasyan\Irefn{triest_i},
R.~Akhunzyanov\Irefn{dubna}, 
M.G.~Alexeev\Irefn{turin_u},
G.D.~Alexeev\Irefn{dubna}, 
A.~Amoroso\Irefnn{turin_u}{turin_i},
V.~Andrieux\Irefnn{illinois}{saclay},
N.V.~Anfimov\Irefn{dubna}, 
V.~Anosov\Irefn{dubna}, 
A.~Antoshkin\Irefn{dubna}, 
K.~Augsten\Irefnn{dubna}{praguectu}
% ...
\end{flushleft}
%
%
\begin{Authlist}
\item \Idef{aveiro}{University of Aveiro, Dept.\ of Physics, 3810-193 Aveiro, Portugal}
\item \Idef{bochum}{Universit\"at Bochum, Institut f\"ur Experimentalphysik, 44780 Bochum, Germany\Arefs{l}\Aref{s}}
\item \Idef{bonniskp}{Universit\"at Bonn, Helmholtz-Institut f\"ur  Strahlen- und Kernphysik, 53115 Bonn, Germany\Arefs{l}}
\item \Idef{bonnpi}{Universit\"at Bonn, Physikalisches Institut, 53115 Bonn, Germany\Arefs{l}}
\item \Idef{brno}{Institute of Scientific Instruments, AS CR, 61264 Brno, Czech Republic\Arefs{m}}
\item \Idef{calcutta}{Matrivani Institute of Experimental Research \& Education, Calcutta-700 030, India\Arefs{n}}
\item \Idef{dubna}{Joint Institute for Nuclear Research, 141980 Dubna, Moscow region, Russia\Arefs{o}}
\item \Idef{erlangen}{Universit\"at Erlangen--N\"urnberg, Physikalisches Institut, 91054 Erlangen, Germany\Arefs{l}}
\item \Idef{freiburg}{Universit\"at Freiburg, Physikalisches Institut, 79104 Freiburg, Germany\Arefs{l}\Aref{s}}
\item \Idef{cern}{CERN, 1211 Geneva 23, Switzerland}
\item \Idef{liberec}{Technical University in Liberec, 46117 Liberec, Czech Republic\Arefs{m}}
\item \Idef{lisbon}{LIP, 1000-149 Lisbon, Portugal\Arefs{p}}
\item \Idef{mainz}{Universit\"at Mainz, Institut f\"ur Kernphysik, 55099 Mainz, Germany\Arefs{l}}
\item \Idef{miyazaki}{University of Miyazaki, Miyazaki 889-2192, Japan\Arefs{q}}
\item \Idef{moscowlpi}{Lebedev Physical Institute, 119991 Moscow, Russia}
\item \Idef{munichtu}{Technische Universit\"at M\"unchen, Physik Dept., 85748 Garching, Germany\Arefs{l}\Aref{r}}
\item \Idef{nagoya}{Nagoya University, 464 Nagoya, Japan\Arefs{q}}
\item \Idef{praguecu}{Charles University in Prague, Faculty of Mathematics and Physics, 18000 Prague, Czech Republic\Arefs{m}}
\item \Idef{praguectu}{Czech Technical University in Prague, 16636 Prague, Czech Republic\Arefs{m}}
\item \Idef{protvino}{State Scientific Center Institute for High Energy Physics of National Research Center `Kurchatov Institute', 142281 Protvino, Russia}
\item \Idef{saclay}{IRFU, CEA, Universit\'e Paris-Saclay, 91191 Gif-sur-Yvette, France\Arefs{s}}
\item \Idef{taipei}{Academia Sinica, Institute of Physics, Taipei 11529, Taiwan\Arefs{tw}}
\item \Idef{telaviv}{Tel Aviv University, School of Physics and Astronomy, 69978 Tel Aviv, Israel\Arefs{t}}
\item \Idef{triest_u}{University of Trieste, Dept.\ of Physics, 34127 Trieste, Italy}
\item \Idef{triest_i}{Trieste Section of INFN, 34127 Trieste, Italy}
\item \Idef{triest_ictp}{Abdus Salam ICTP, 34151 Trieste, Italy}
\item \Idef{turin_u}{University of Turin, Dept.\ of Physics, 10125 Turin, Italy}
\item \Idef{turin_i}{Torino Section of INFN, 10125 Turin, Italy}
\item \Idef{illinois}{University of Illinois at Urbana-Champaign, Dept.\ of Physics, Urbana, IL 61801-3080, USA\Arefs{nsf}}
\item \Idef{warsaw}{National Centre for Nuclear Research, 00-681 Warsaw, Poland\Arefs{u}}
\item \Idef{warsawu}{University of Warsaw, Faculty of Physics, 02-093 Warsaw, Poland\Arefs{u}}
\item \Idef{warsawtu}{Warsaw University of Technology, Institute of Radioelectronics, 00-665 Warsaw, Poland\Arefs{u} }
\item \Idef{yamagata}{Yamagata University, Yamagata 992-8510, Japan\Arefs{q} }
\end{Authlist}
%
%
\renewcommand\theenumi{\alph{enumi}}
\begin{Authlist}
\item [{\makebox[2mm][l]{\textsuperscript{\#}}}] Corresponding authors
\item [{\makebox[2mm][l]{\textsuperscript{*}}}] Deceased
\item \Adef{a}{Also at Instituto Superior T\'ecnico, Universidade de Lisboa, Lisbon, Portugal}
\item \Adef{b}{Also at Dept.\ of Physics, Pusan National University, Busan 609-735, Republic of Korea and at Physics Dept., Brookhaven National Laboratory, Upton, NY 11973, USA}
\item \Adef{r}{Supported by the DFG cluster of excellence `Origin and Structure of the Universe' (www.universe-cluster.de) (Germany)}
\item \Adef{p2i}{Supported by the Laboratoire d'excellence P2IO (France)}
\item \Adef{d}{Also at Chubu University, Kasugai, Aichi 487-8501, Japan\Arefs{q}}
\item \Adef{x}{Also at Dept.\ of Physics, National Central University, 300 Jhongda Road, Jhongli 32001, Taiwan}
\item \Adef{e}{Also at KEK, 1-1 Oho, Tsukuba, Ibaraki 305-0801, Japan}
\item \Adef{g}{Also at Moscow Institute of Physics and Technology, Moscow Region, 141700, Russia}
\item \Adef{v}{Supported by Presidential Grant NSh--999.2014.2 (Russia)}
\item \Adef{h}{Present address: RWTH Aachen University, III.\ Physikalisches Institut, 52056 Aachen, Germany}
\item \Adef{y}{Also at Dept.\ of Physics, National Kaohsiung Normal University, Kaohsiung County 824, Taiwan}
\item \Adef{turin_p}{Also at University of Eastern Piedmont, 15100 Alessandria, Italy}
\item \Adef{i}{Present address: Uppsala University, Box 516, 75120 Uppsala, Sweden}
\item \Adef{c}{  Supported by the DFG Research Training Group Programmes 1102 and 2044 (Germany)} 
%
%
\item \Adef{l}{  Supported by BMBF - Bundesministerium f\"ur Bildung und Forschung (Germany)}
\item \Adef{s}{  Supported by FP7, HadronPhysics3, Grant 283286 (European Union)}
\item \Adef{m}{  Supported by MEYS, Grant LG13031 (Czech Republic)}
\item \Adef{n}{  Supported by SAIL (CSR) and B.Sen fund (India)}
\item \Adef{o}{  Supported by CERN-RFBR Grant 12-02-91500}
\item \Adef{p}{\raggedright 
                 Supported by FCT - Funda\c{c}\~{a}o para a Ci\^{e}ncia e Tecnologia, COMPETE and QREN, Grants CERN/FP 116376/2010, 123600/2011 
                 and CERN/FIS-NUC/0017/2015 (Portugal)}
\item \Adef{q}{Supported by MEXT and JSPS, Grants 18002006, 20540299, 18540281 and  
26247032, the Daiko and Yamada Foundations (Japan) }
\item \Adef{tw}{ Supported by the Ministry of Science and Technology (Taiwan)}
\item \Adef{t}{Supported by the Israel Academy of Sciences and Humanities (Israel)}
\item \Adef{nsf}{Supported by the National Science Foundation under grant no. PHY-1506416 (USA)}
\item \Adef{u}{  Supported by NCN, Grant 2015/18/M/ST2/00550 (Poland)}
\end{Authlist}
%
%
%
%
%

\end{titlepage}

According to Quantum Chromodynamics (QCD), the modern theory of strong interactions, the internal structure of hadrons explored in hard (semi-)inclusive scattering is described by parton distribution functions (PDFs). For a polarized nucleon, within the twist-2 approximation there are eight transverse-momentum-dependent (TMD) PDFs describing the distributions of longitudinal and transverse momenta of partons and their correlations with nucleon and quark polarizations. Experimentally, these PDFs can be accessed in the Drell-Yan process, \textit{i.e.} massive lepton-pair production in hadron-nucleon collisions ($h \, N \rightarrow \ell\,\bar{\ell}\, X$), hereafter referred to as DY, and in semi-inclusive hadron measurements in deep-inelastic lepton-nucleon scattering ($\ell \,N \rightarrow \ell^\prime \,h \, X$), hereafter referred to as SIDIS. For recent reviews see \textit{e.g.} Refs.~\cite{Perdekamp:2015vwa,Peng:2014hta,Aidala:2012mv}.
For the DY and SIDIS cross sections, TMD factorization was proven to hold~\cite{Collins:2011zzd}, which allows one to express them as convolutions of hard-scale dependent TMD PDFs, perturbatively calculable parton hard-scattering cross sections and  (for SIDIS) parton fragmentation functions. The hard scale $Q$ in DY is given by the invariant mass of the lepton pair and in SIDIS by the square root of the virtuality of the photon exchanged in the DIS process.

The Sivers function~\cite{Sivers:1989cc} plays an important role among the TMD PDFs. It describes the left-right asymmetry in the distribution of unpolarized partons in the nucleon with respect to the plane spanned by the momentum and spin vectors of the nucleon.
One of the recent significant theoretical advances in the TMD framework of QCD is the prediction that the two naively time-reversal odd TMD PDFs, \textit{i.e.} the quark Sivers functions $f_{1T}^\perp$ and Boer-Mulders functions $h_{1}^\perp$, have opposite sign when measured in SIDIS or DY
~\cite{Collins:2002kn, Brodsky:2002cx, Brodsky:2002rv}.
The experimental test of this fundamental prediction, which is a direct consequence of QCD gauge invariance, is a major challenge in hadron physics.

Non-zero quark Sivers TMD PDFs have been extracted from SIDIS single-differential results of HERMES~\cite{Airapetian:2009ae}, COMPASS~\cite{Ageev:2006da,Alekseev:2008aa, Adolph:2012sp,Adolph:2014zba} and JLab~\cite{Qian:2011py} using both collinear~\cite{Anselmino:2005ea, Anselmino:2008sga} and TMD evolution approaches~\cite{Aybat:2011ta, Anselmino:2012aa, Anselmino:2016uie,Echevarria:2014xaa,Sun:2013hua}.
%
The first measurement of the Sivers effect in $W$ and $Z$-boson production in collisions of transversely polarized protons at RHIC was reported by the STAR collaboration~\cite{Adamczyk:2015gyk};
the hard scales of these measurements is $Q\approx 80$\;\gvc\; and $90$\;\gvc. It is quite different from the one explored in fixed-target experiments where $Q$ ranges approximately between $1$\;\gvc\; and $9$\;\gvc. Hence it is not excluded that TMD evolution effects may play a substantial role when describing the STAR results using Sivers TMD PDFs extracted from fixed-target SIDIS results.

The COMPASS experiment at CERN~\cite{Abbon:2007pq, Gautheron:2010wva} has the unique capability to explore the transverse-spin structure of the nucleon in a similar kinematic region by two alternative experimental approaches, \textit{i.e.} SIDIS and DY, using mostly the same setup.
This offers the opportunity of minimizing uncertainties of TMD evolution in the comparison of the Sivers TMD PDFs when extracted from these two measurements to test the opposite-sign prediction by QCD.

Recently, COMPASS published the first multi-differential results of the Sivers asymmetry, which were extracted from SIDIS data at four different hard scales~\cite{Adolph:2016dvl}. In particular for the range 4 \gvc$\;<Q<$ 9 \gvc, the Sivers asymmetry for positive hadrons was found to be above zero by 3.2 standard deviations of the total experimental accuracy. This hard scale range is very similar to the one used in this Letter to analyze the DY process.

\begin{figure}[bp]
\centering
\includegraphics[width=0.35\textwidth]{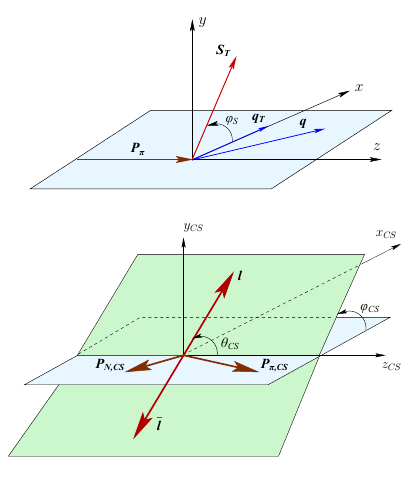}
\caption{Top: target rest frame.
Bottom: Collins-Soper frame.
%
}
\label{fig:DYframe}
\end{figure}

When the polarizations of the produced leptons are summed over, the general expression for the cross section of pion-nucleon DY lepton-pair production off a transversely polarized nucleon comprises five transverse spin-dependent asymmetries (TSAs), including the Sivers TSA~\cite{Arnold:2008kf,Gautheron:2010wva}. Those three TSAs that can be described by contributions from only twist-2 TMD PDFs will be addressed in this Letter. The corresponding part of the differential cross section  can be written as follows:
\vskip 0.5cm
\begin{eqnarray}\label{eq:DY_xsecLO}\nonumber
  \hspace*{-0.9cm}\frac{d\sigma}{dq^{4}d\Omega} &\propto& \hat{\sigma}_{U}
  \bigg\{ 1 +  {S_T}\Big[ D_{1}A_T^{\sin\phiS}\sin\phiS\\ \nonumber
   &&
       +D_{2}\Big( A_T^{\sin(2\phiCS-\phiS)}\sin(2\phiCS-\phiS) \\
    &&
   \qquad+ A_T^{\sin(2\phiCS+\phiS)}\sin(2\phiCS+\phiS) \Big)
   \Big] \bigg\}.
\end{eqnarray}
%
Here, $q$ is the four-momentum of the exchanged virtual photon and $\hat{\sigma}_{U} =  \lf({F^{1}_{U}}+{F^{2}_{U}}\rg)\lf(1 + \lambda{{\cos}^2}\thCS \rg)$, with $F^{1}_{U}$, $F^{2}_{U}$ being the polarization and azimuth-independent structure functions, and the polar asymmetry $\lambda$ is given as $\lambda=\lf({F^{1}_{U}}-{F^{2}_{U}}\rg)/\lf({F^{1}_{U}}+{F^{2}_{U}}\rg)$. At leading order of  perturbative QCD, within the twist-2 approximation, $F^{2}_{U}=0$ and therefore $\lambda=1$. The subscript ($U$)$T$ denotes transverse polarization (in)dependence. In analogy to SIDIS, the virtual-photon depolarization factors are defined as $D_{1}=(1+{{\cos}^2}\thCS)/\lf(1 + \lambda{{\cos }^2}\thCS \rg)$ and $D_{2}={{\sin}^2}\thCS/\lf(1 + \lambda{{\cos }^2}\thCS \rg)$. The angles $\phiCS$, $\thCS$ and $\Omega$, the solid angle of the lepton, are defined in the Collins-Soper frame as defined in Refs.~\cite{Arnold:2008kf,Gautheron:2010wva}, and $\phiS$ is the azimuthal angle of the direction of the nucleon polarization in the target rest frame, see Fig.~\ref{fig:DYframe}.
%
%
%
%
%

The TSAs $A_{T}^{w}$ in Eq.~(\ref{eq:DY_xsecLO}) are defined as amplitudes of a given azimuthal modulation $w=w(\phiS,\phiCS)$, divided by the spin and azimuth-independent part of the DY cross section and the corresponding depolarization factor.

In this analysis, the sign convention for TSAs is given by Eq.~(\ref{eq:DY_xsecLO}) together with the definitions of azimuthal and polar angles in Fig.~\ref{fig:DYframe}.
Note that the sign of the Sivers TSA is related to that of the Sivers TMD PDF only by convention, so that the above mentioned sign-change prediction for the Sivers TMD PDFs in our case results in the same sign of the measured Sivers TSAs in SIDIS and DY.

In DY lepton-pair production with a transversely polarized nucleon in the initial state, the TSA $A_T^{\sin\phiS}$ is related to the nucleon Sivers TMD PDFs ($f_{1T}^\perp$) convoluted with the unpolarized pion TMD PDFs ($f_{1,\pi}$). The other two TSAs, $A_T^{\sin(2\phiCS-\phiS)}$ and $A_T^{\sin(2\phiCS+\phiS)}$, are related to convolutions of the Boer-Mulders TMD PDFs ($h_{1,\pi}^{\perp}$) of the pion with the nucleon TMD PDFs transversity ($h_1$) and pretzelosity ($h_{1T}^\perp$), respectively~\cite{Bacchetta:2006tn,Arnold:2008kf}.
All three aforementioned nucleon TMD PDFs induce analogous twist-2 TSAs in the general expression for the cross section of unpolarized-hadron production in SIDIS of leptons off transversely polarized nucleons~\cite{Kotzinian:1994dv,Bacchetta:2006tn,Arnold:2008kf}. These TSAs were studied by the HERMES and COMPASS experiments~\cite{Alekseev:2008aa,Airapetian:2009ae,Adolph:2012sp,Adolph:2014fjw,Adolph:2014zba,Adolph:2016dvl,Adolph:2012sn,Adolph:2012nw,Parsamyan:2013ug,Pappalardo:2010zz,Bradamante:2017yia}.
In contrast to the  Sivers function, transversity and pretzelosity are predicted to be genuinely universal, \textit{i.e.} they do not change sign between SIDIS and DY~\cite{Collins:2011zzd}, which is yet another fundamental QCD prediction to be explored.


%
The analysis presented in this Letter is based on Drell-Yan data collected by COMPASS in the year 2015 using essentially the same spectrometer as it was used during SIDIS data taking in previous years~\cite{Abbon:2007pq}. For this measurement, the 190 \gvc\; $\pi^{-}$ beam with an average intensity of $0.6\times10^{8}$ s$^{-1}$ from the CERN SPS was scattered off the COMPASS transversely polarized NH$_3$ target with proton polarization $\langle P_T\rangle\approx0.73$ and
dilution factor $\langle f\rangle\approx 0.18$, where the latter accounts for the fraction of polarizable nucleons in the target and the migration of reconstructed events from one target cell to the other.
The polarized target, placed in a 0.6~T dipole magnet, consisted of two longitudinally aligned cylindrical cells of 55 cm length and 4~cm in diameter, separated
by a 20~cm gap.
The two cells were polarized vertically in opposite directions, so that data with both spin orientations were recorded simultaneously. In order to compensate for acceptance effects, the polarization was reversed every two weeks.
The entire data-taking time of 18 weeks was divided into nine periods, each consisting of two consecutive weeks with opposite target polarizations.
The proton polarization had a relaxation time of about 1000 hours, which was measured for each target cell in each data taking period.
A 240 cm long structure made mostly of alumina with a tungsten core, placed downstream of the target, acted as hadron absorber and beam dump.
Outgoing charged particles were detected by a system of tracking detectors in the two-stage spectrometer. In each stage, muon identification was accomplished by a system of muon filters.

The trigger required the hit pattern of
several hodoscope planes to be consistent with at least two muon candidates originating from the target region. For any pair of candidates either both have to be detected in the first stage of the spectrometer ($25 < \theta_{\mu} < 160$~mrad), or one in the first and the other in the second stage ($8 < \theta_{\mu} < 45$~mrad).

%
%
%

In the data analysis, the selection of events requires a production vertex located within the polarized-target volume, with one incoming pion beam track and at least two oppositely charged outgoing particles that are consistent with the muon hypothesis, \textit{i.e.} they crossed at least 30 radiation lengths of material along the spectrometer.
%
%
The dimuon transverse momentum $q_T$ is required to be above $0.4$ \gvc\; in order to obtain sufficient resolution in angular variables.

\begin{figure}[tbp]
 \begin{minipage}[t]{0.47\linewidth}
 \center
 \includegraphics[width=0.9\textwidth]{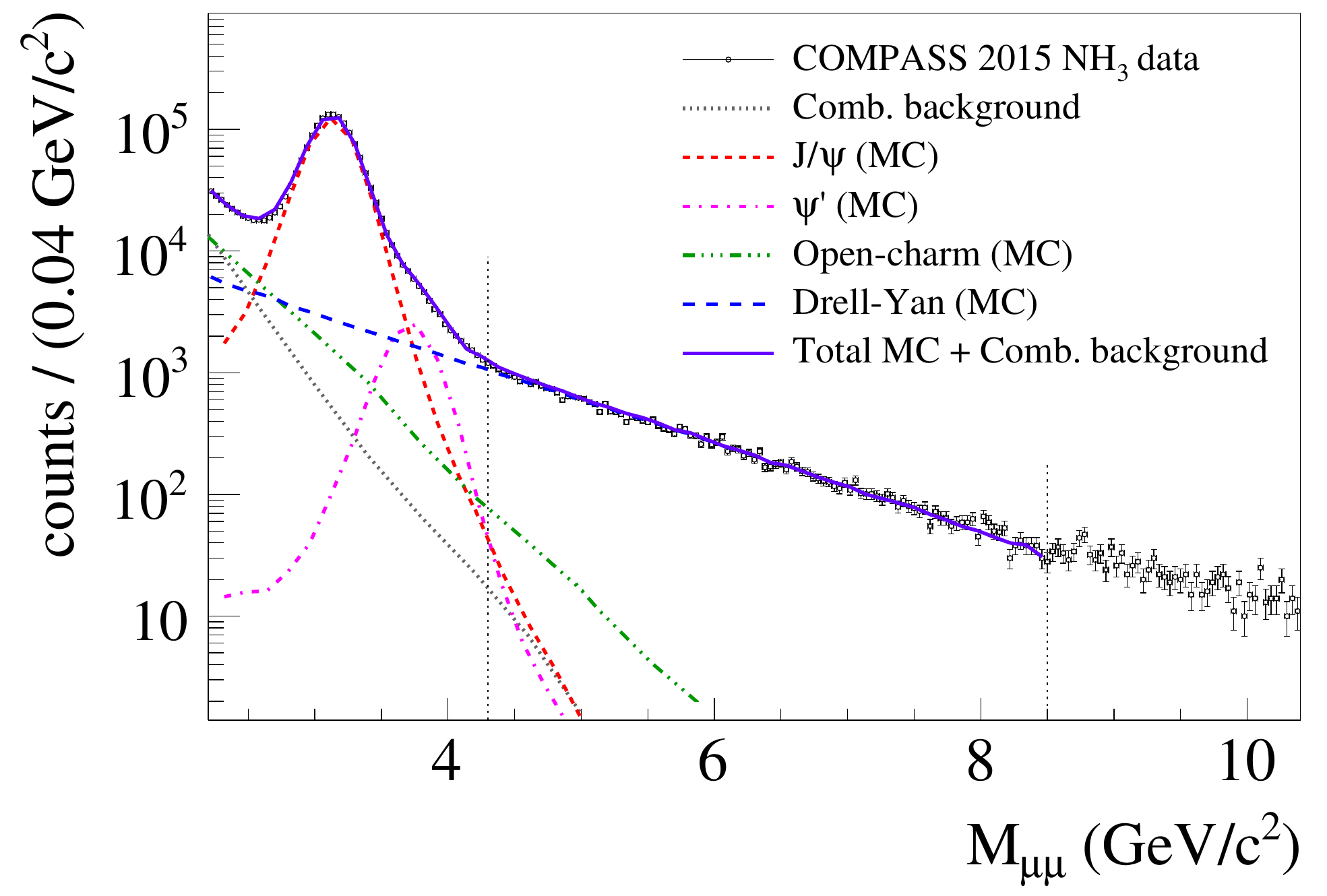}
 \caption{The dimuon invariant mass distribution.}
 \label{fig:M}
 \end{minipage}
 \begin{minipage}[t]{0.53\linewidth}
 \center
 \includegraphics[width=0.9\textwidth]{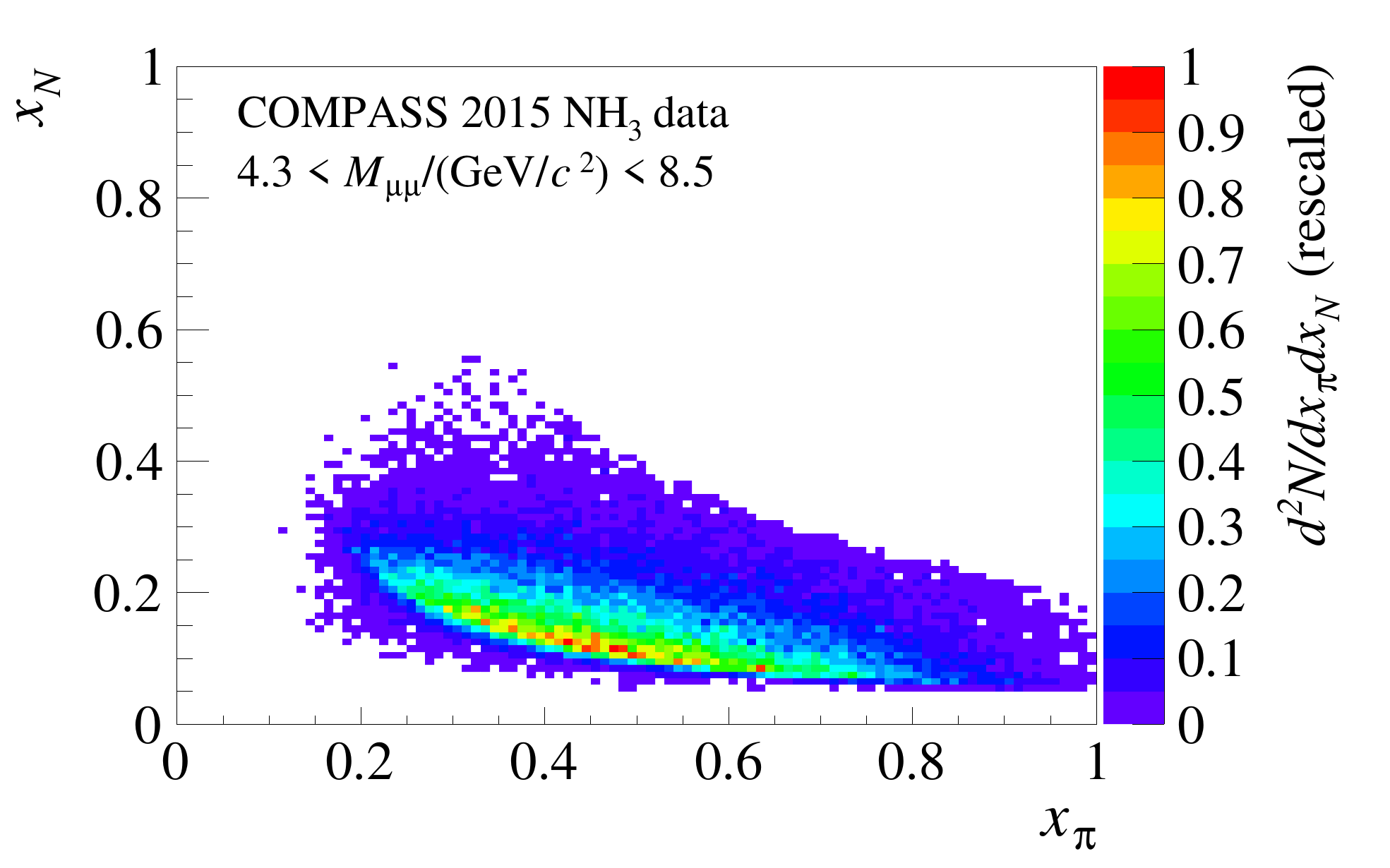}
 \caption{The two-dimensional $(x_{\pi},x_{N})$ distribution of the selected high mass dimuons. The distribution is normalized to have a maximum value equal to one.}
 \label{fig:x1x2}
 \end{minipage}
\end{figure}

%

The reconstructed mass spectrum of events passing all analysis requirements is shown in Fig.~\ref{fig:M} (in black). The J/$\psi$ peak is clearly visible with a shoulder
from the $\psi$(2S) resonance.  The contributions to the dimuon spectrum are evaluated with a Monte-Carlo (MC) simulation. The Drell-Yan process is shown in blue (long dashed) together with background processes: i) charmonia in red (dashed) and magenta (dot-dashed) and ii) semi-muonic open-charm decays shown in green (double dot dashed).
In addition, the combinatorial background originating from the decays of pions and kaons produced in the target is evaluated using like-sign dimuon events from real data and shown in grey (dotted).
The sum of all contributions, shown in violet, describes the
experimental data well. For the analysis we use the mass range 4.3~\gvcw\,$ < M_{\mu\mu} < $\,8.5~\gvcw\,, where the upper limit avoids the contribution of $\Upsilon$-resonances. In this range, the sum of all background contributions is estimated to be below 4\%.

The two-dimensional distribution of the Bjorken scaling variables of pion and nucleon, $x_{\pi}$ and $x_{N}$, for this mass range is presented in Fig.~\ref{fig:x1x2}. The figure shows that the kinematic phase space explored by the COMPASS spectrometer matches the valence region in $x_{\pi}$ and $x_{N}$. In this region, the DY cross section for a proton target is dominated by the contribution of nucleon $u$-quark and pion $\bar{u}$-quark TMD PDFs.

\begin{figure}[tbp]
\centering
\includegraphics[width=0.40\textwidth]{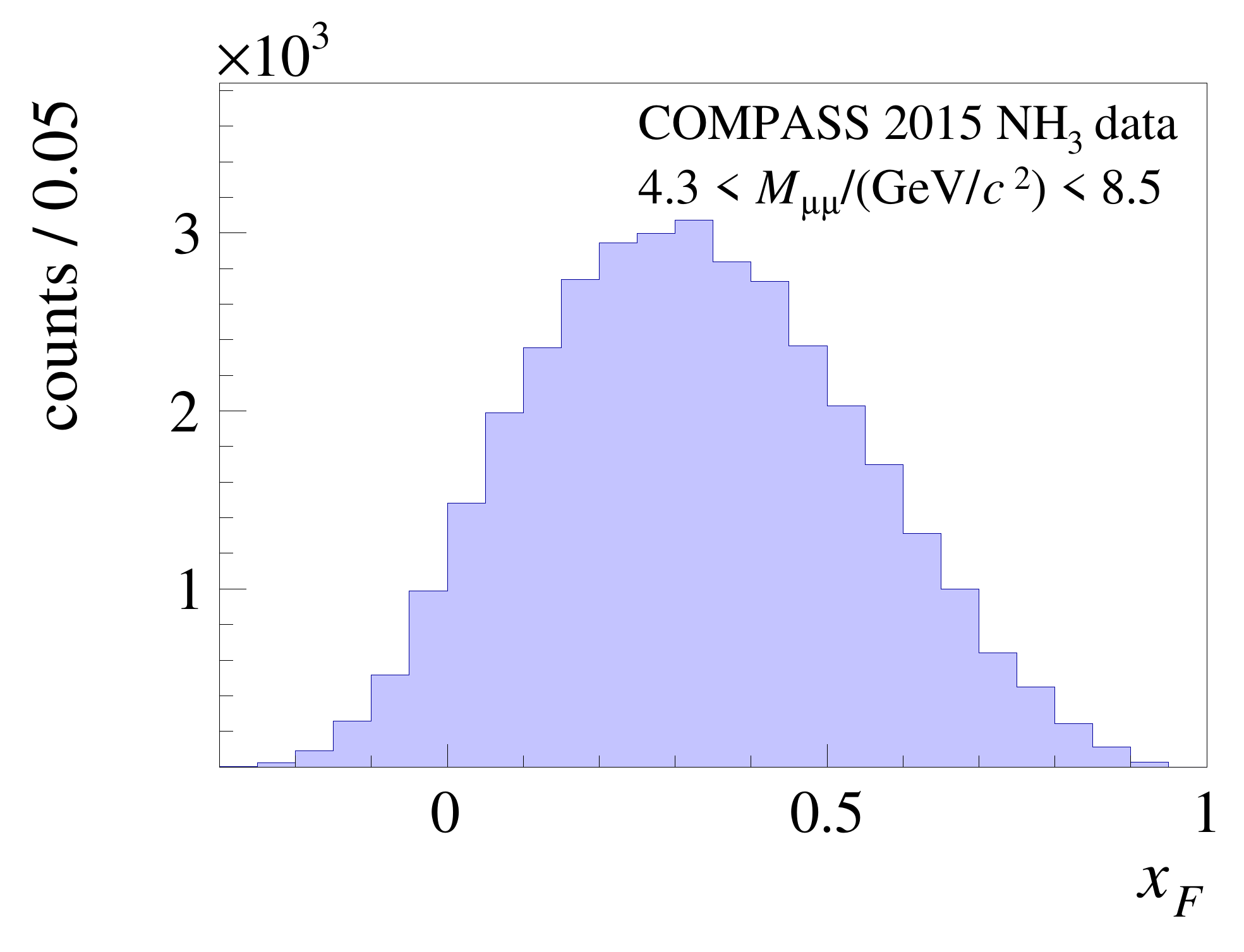}
\hskip 1cm
\includegraphics[width=0.40\textwidth]{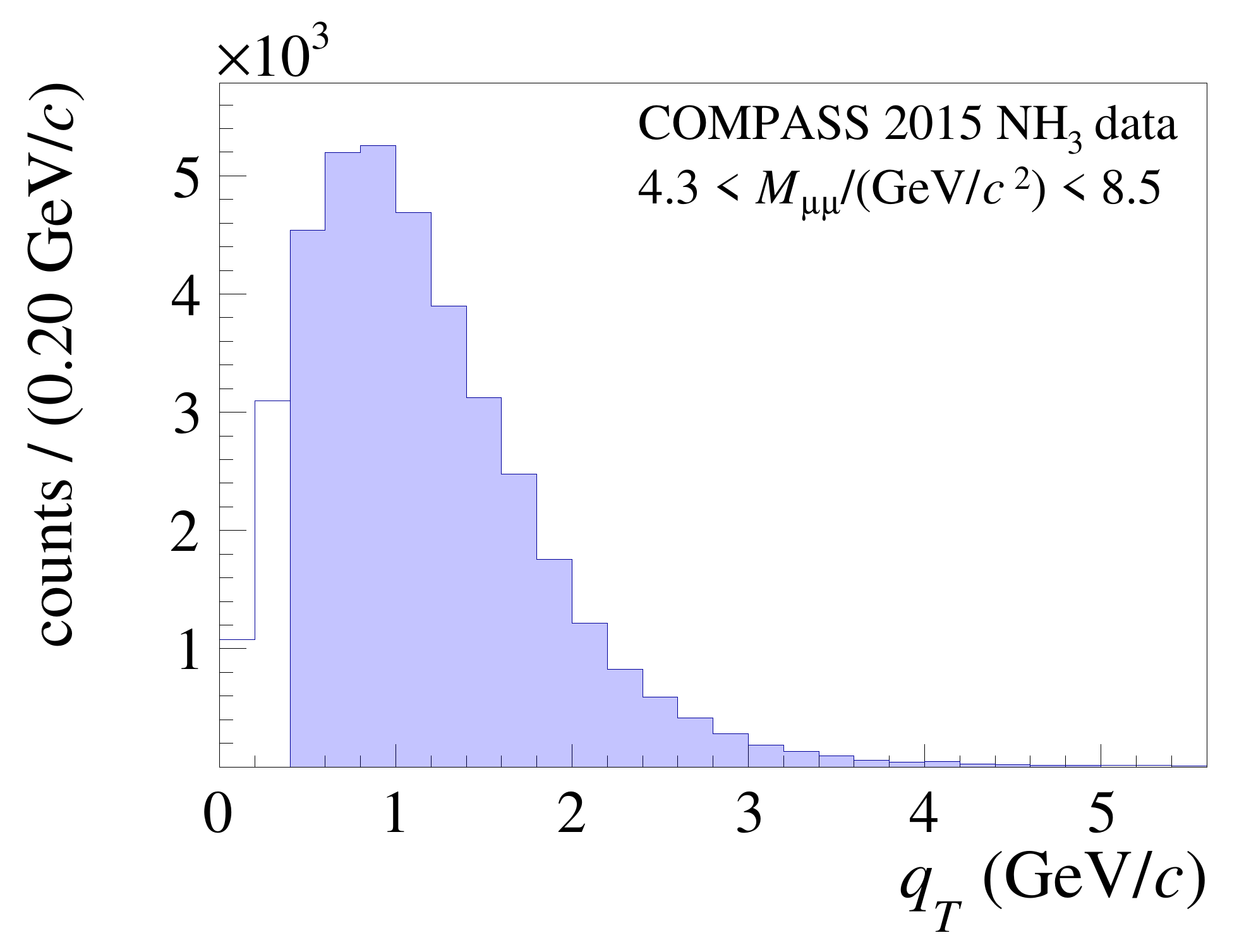}
\caption{The $x_{F}$ distribution (left) and $q_T$ distribution (right)
of the selected high mass dimuons.}
\label{fig:xF_qT}
\end{figure}

The distributions of the dimuon Feynman variable $x_{F}$ and the dimuon transverse momentum $q_{T}$ are presented in Fig.~\ref{fig:xF_qT}. The corresponding mean values of the kinematic variables are: $\langle x_{N} \rangle=0.17$,
$\langle x_{\pi} \rangle=0.50$,
$\langle x_{F} \rangle=0.33$,
$\langle q_{T} \rangle=1.2$~\gvc\, and
$\langle M_{\mu\mu} \rangle=5.3$~\gvcw.

About $35\times10^3$ dimuons remain for the analysis.
The three TSAs presented in this Letter are extracted period by period from the number of dimuons produced in each cell for the two directions of the target polarization.
The double-cell target configuration in conjunction with the periodic polarization reversal allows for the simultaneous measurement of azimuthal asymmetries for both target spin orientations, thereby compensating flux and acceptance-dependent systematic uncertainties.
Using an extended Unbinned Maximum Likelihood estimator as described in Ref.~\cite{Adolph:2012nw}, all five TSAs are fitted simultaneously together with their correlation matrices.
The final asymmetries are obtained by averaging the results of the nine periods.
The asymmetries are evaluated in kinematic bins of $x_{N}$, $x_{\pi}$, $x_{F}$ or $q_{T}$, while always integrating over
all the other variables.


The dilution factor $f$ and the depolarization factors $D_{1}$ and $D_{2}$ entering the definition of TSAs are calculated on an event-by-event basis and used to weight the asymmetries. For the magnitude of the target polarization $P_T$, an average value is used for each data taking period in order to avoid possible systematic bias. In the evaluation of the depolarization factors, the approximation $\lambda=1$ is used. Known deviations from this assumption with $\lambda$ ranging between 0.5 and 1~\cite{Guanziroli:1987rp,Conway:1989fs} lead to a normalization uncertainty of at most $-5\%$.



The TSAs resulting from different periods are checked for possible systematic effects. The largest systematic uncertainty is due to possible residual variations of experimental conditions within a given period. They are quantified by evaluating various types of false asymmetries in a similar way as described in Refs.~\cite{Adolph:2012sp,Adolph:2012sn}. The systematic point-to-point uncertainties are found to be about 0.7 times the statistical uncertainties.
The normalization uncertainties originating from the uncertainties on target polarization ($5\%$) and dilution factor ($8\%$) are not included in the quoted systematic uncertainties.
%
%
%
%
%
%
\begin{figure}[tbp]
\centering
\includegraphics[width=0.48\textwidth]{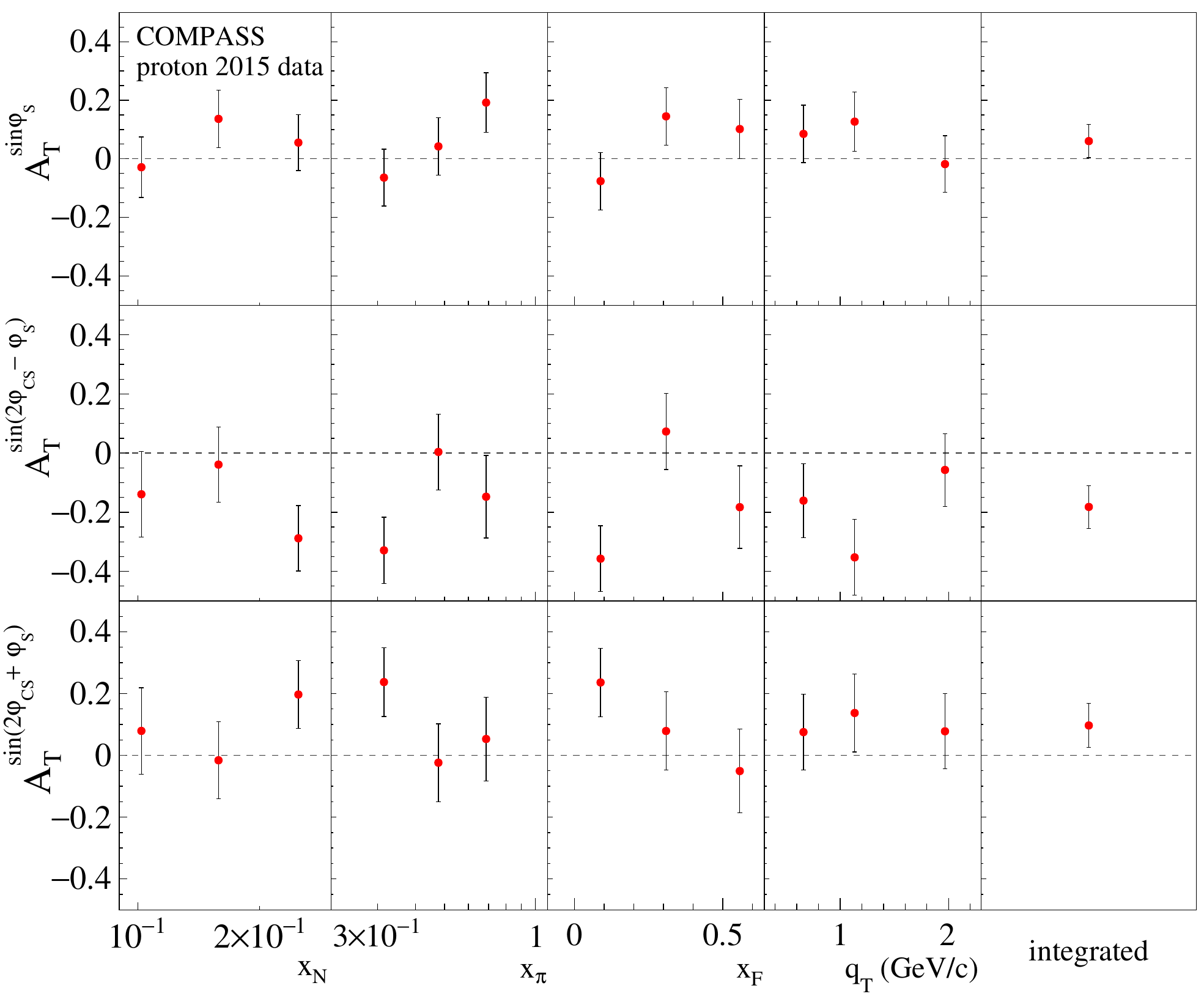}
\caption{Extracted Drell-Yan TSAs related to Sivers, transversity and pretzelosity TMD PDFs (top to bottom). Error bars represent statistical uncertainties. Systematic uncertainties (not shown) are 0.7 times the statistical ones.}
\label{fig:TSAs}
\end{figure}

\begin{figure}[tbp]
\centering
\includegraphics[width=0.39\textwidth]{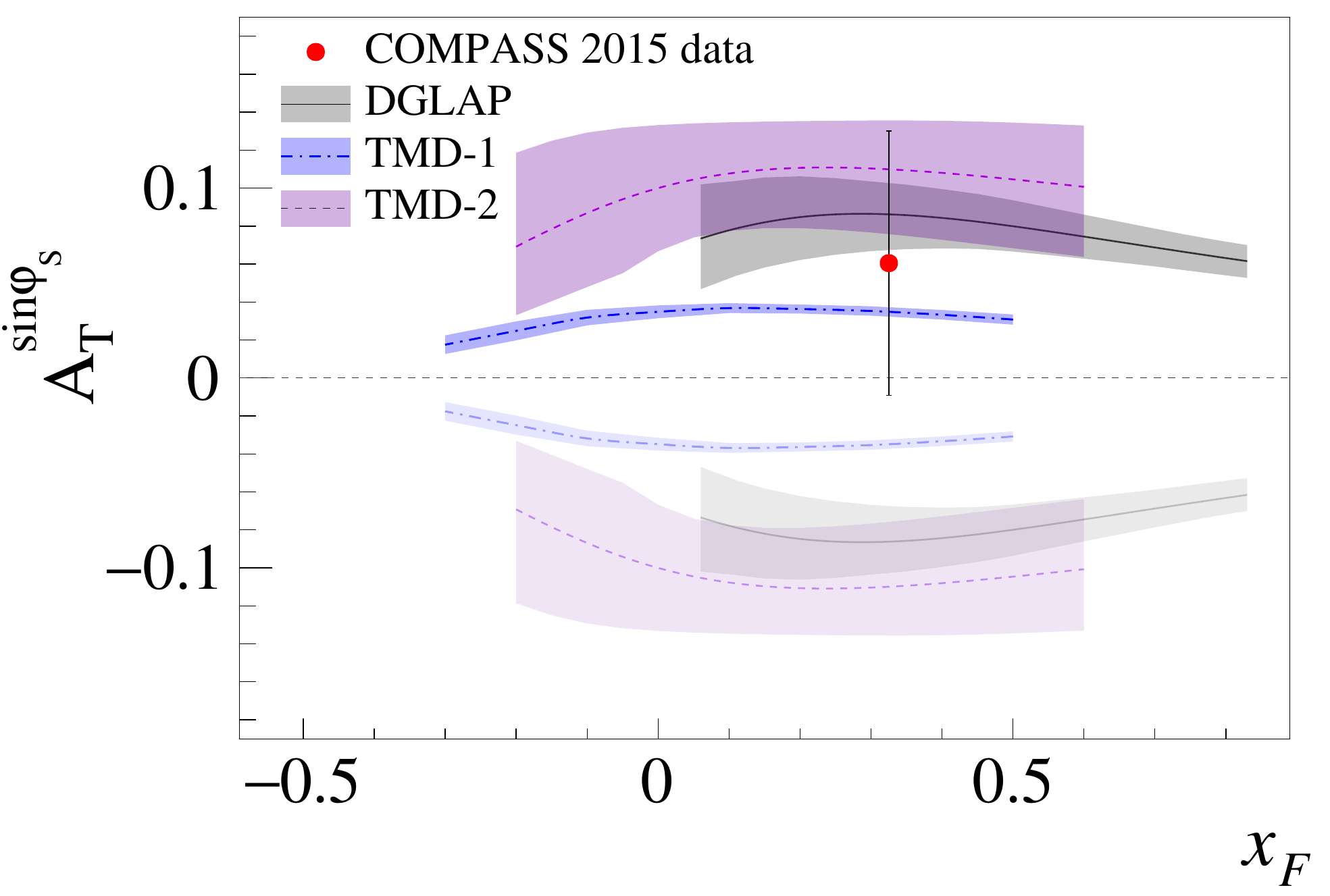}
\caption{The measured mean Sivers asymmetry and the
theoretical  predictions for different $Q^2$ evolution schemes from Refs.~\cite{Anselmino:2016uie} (DGLAP),~\cite{Echevarria:2014xaa} (TMD1)
and~\cite{Sun:2013hua} (TMD2).
The dark-shaded (light-shaded) predictions are evaluated with (without) the sign-change hypothesis. The error bar represents the total experimental uncertainty.}
\label{fig:Siv_theor}
\end{figure}

The TSAs $A_T^{\sin\phiS}$, $A_T^{\sin(2\phiCS-\phiS)}$ and $A_T^{\sin(2\phiCS+\phiS)}$ are shown in Fig.~\ref{fig:TSAs} as a function of the variables $x_{N}$, $x_{\pi}$, $x_{F}$ and $q_{T}$.
Due to relatively large statistical uncertainties, no clear trend is observed for any of the TSAs.
The full set of numerical values for all TSAs including correlation coefficients and mean kinematic values from this measurement is available on HepData~\cite{hepdata}.
%
%
The last column in Fig.~\ref{fig:TSAs} shows the results for the three extracted TSAs integrated over the entire kinematic range.
The average Sivers asymmetry $A_T^{\sin\phiS}$ is found to be above zero at about one standard deviation of the total uncertainty. In Fig.~\ref{fig:Siv_theor}, it is compared with recent theoretical predictions from Refs.~\cite{Anselmino:2016uie,Echevarria:2014xaa,Sun:2013hua} that are based on different $Q^2$-evolution approaches. The positive sign of these theoretical predictions for the DY Sivers asymmetry was obtained by using the sign-change hypothesis for the Sivers TMD PDFs, and the numerical values are based on a fit of SIDIS data for the Sivers TSA~\cite{Airapetian:2009ae,Alekseev:2008aa, Adolph:2012sp}.
The figure shows that this first measurement of the DY Sivers asymmetry is consistent with the predicted change of sign for the Sivers function.
%

%
%
%
%
%
%

The average value for the TSA  $A_T^{\sin(2\phiCS-\phiS)}$ is measured to be below zero with a significance of about two standard deviations. The obtained magnitude of the asymmetry is in agreement with the model calculations of Ref.~\cite{Sissakian:2010zza} and can be used to study the universality of the nucleon transversity function. The TSA $A_T^{\sin(2\phiCS+\phiS)}$, which is related to the nucleon pretzelosity TMD PDFs, is measured to be above zero with a significance of about one standard deviation. Since both $A_T^{\sin(2\phiCS-\phiS)}$ and $A_T^{\sin(2\phiCS+\phiS)}$ are related to the pion Boer-Mulders PDFs, the obtained results may be used to study this function further and to possibly determine its sign.
They may also be used to test the sign change of the nucleon Boer-Mulders TMD PDFs between SIDIS and DY as predicted by QCD~\cite{Collins:2002kn, Brodsky:2002cx, Brodsky:2002rv}, when combined with other past and future SIDIS and DY data related to target-spin-independent Boer-Mulders asymmetries~\cite{Falciano:1986wk,Airapetian:2012yg,Adolph:2014pwc}.
%

%



%
%

We gratefully acknowledge the support of the CERN management and staff and the skill and effort of the technicians of our collaborating institutes. Special thanks go to A. Dudarev, E. Feldbaumer, L. Gatignon, C. Theis, H. ten Kate and H. Vincke for invaluable help in the preparation of experiment.
For fruitful discussions and essential input to the preparation of the proposal of this measurement we thank M. Anselmino, A. Bacchetta, A. Bianconi, S. Melis, M Radici, O. Teryaev, B. Pasquini and A. Prokudin.
We are grateful to M.E.~Boglione, M.~Echevarria and F.~Yuan for providing us with numerical values of their model predictions.


%
%
%

\bibliography{DY_paper}{}

\end{document}